\documentclass[aps,prb,showpacs,twocolumn,superscriptaddress]{revtex4-2}

\usepackage{graphicx} 
\usepackage{color}
\usepackage{array}
\usepackage{mathrsfs}
\usepackage{ulem} 
\usepackage{amsmath}

\begin{document}
\title{Isothermal and adiabatic magnetization processes of the spin-$\frac{1}{2}$ Heisenberg model on an anisotropic triangular lattice}

\author{Katsuhiro Morita}
\email[e-mail:]{katsuhiro.morita@rs.tus.ac.jp}
\affiliation{Department of Physics, Faculty of Science and Technology, Tokyo University of Science, Chiba 278-8510, Japan}

\begin{abstract}
 In this study, we investigate the magnetic susceptibility, entropy, and isothermal magnetization curve of the spin-1/2 Heisenberg model on an anisotropic triangular lattice using the orthogonalized finite-temperature Lanczos method.
In addition, we investigate the adiabatic magnetization curve and magnetocaloric effect.
We estimate these physical quantities with sufficient accuracy in the thermodynamic limit, except at low temperatures.
We observe a 1/3 magnetization plateau in the isothermal magnetization process, whereas the plateau is observed to have a slope in the adiabatic process.
We show that the magnetocaloric effect can be used to detect the signature of phase transitions.
We believe that these results will be useful for understanding the magnetism of anisotropic triangular lattice compounds through a comparison with experimental results in the future.
\end{abstract}

\maketitle
\section{Introduction}

Megagauss magnetic field generators have been developed for over half a century~\cite{revew1,revew2,T1,T2,T3,T4}. 
In recent years, the magnetization process of magnetic materials has been actively studied using megagauss magnetic field generators, and various quantum phase transitions have been successfully observed~\cite{E1,E2,E3,E4,E5,E6,E7,E8,E9}. 
Most experiments using magnetic fields exceeding 100 T have been performed with pulse widths of a few to several tens of microseconds~ \cite{revew1,revew2}.
Owing to the very narrow pulse width, the magnetization process is expected to be an adiabatic (isentropic) process rather than an isothermal process. 
In addition, several studies have observed the magnetocaloric effect in magnetic compounds~\cite{E3,E6,MCE1,MCE2,MCE3}. 
Therefore, a theoretical study of the adiabatic magnetization process is important for understanding these experimental results.

The spin-1/2 Heisenberg model on isotropic and anisotropic triangular lattices is a traditional model used in studies on magnetism.
This model has been extensively investigated for several decades~\cite{ITLrev}.
Several model compounds have been studied, and their magnetization processes at low temperatures have been observed to exhibit a 1/3 magnetization plateau and various phase transitions because of the frustration and quantum effects~\cite{CCC1,CCC2,CCC3,CCB1,CCB2,CCCB,Cotri1,Cotri2,Cotri3}.
In addition, in theoretical studies, magnetic-field-induced quantum phase transitions at zero temperature have been found in the Heisenberg model on an anisotropic triangular lattice~\cite{ITLthe,ITLSWT,ITLthe1}. 

Very recently, $A_3$ReO$_5$Cl$_2$ ($A$ = Ca, Ba, Sr) with the spin-1/2 anisotropic triangular lattice have been intensively studied~\cite{Re1,Re2,Re3}.
Therefore, theoretical calculations of the isothermal and adiabatic magnetization curves and magnetocaloric effect in the anisotropic triangular lattice Heisenberg model are necessary for future experimental studies. 
However, the calculations of the adiabatic magnetization process for frustrated spin-1/2 systems are limited to approximately 20 sites using full exact diagonalization (FullED)~\cite{MCEthe}.
Thus, numerical calculations with a larger size are necessary to estimate the physical quantities of the anisotropic triangular lattice compounds.

\begin{figure}[tb]
\includegraphics[width=86mm]{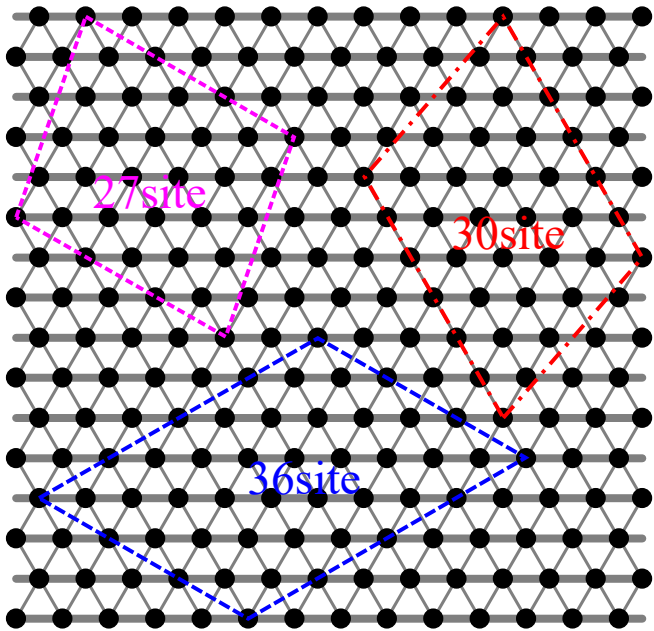}
\caption{Lattice structure of the anisotropic triangular lattice with exchange interactions, $J$ and $J'$. 
The solid and thin lines represent $J$ and $J'$, respectively. We set $J=1$.
The black circles represent the sites with a spin.
The pink, red, and blue dashed quadrangles represent the clusters of $N = 27$, $N = 30$, and $N = 36$, respectively, used in the OFTLM with periodic boundary conditions, where $N$ is the number of sites.
\label{lattice}}
\end{figure}

In this study, we investigate the magnetic properties of the spin-1/2 anisotropic triangular lattice with exchange interactions $J$ and $J'$, as shown in Fig.~\ref{lattice} using the orthogonalized finite-temperature Lanczos method (OFTLM)~\cite{OFTL}, which is an improved version of the standard finite-temperature Lanczos method (FTLM)~\cite{ftl1,ftl2}.
 The OFTLM can be used to evaluate the adiabatic process with high accuracy because the value of the entropy is almost exact at low temperatures, whereas the entropy calculated by the standard FTLM has a somewhat large standard error at low temperatures without sufficient sampling over random vector~\cite{OFTL}.
We first investigate the magnetic susceptibility and magnetic entropy of the anisotropic triangular lattice up to 36 sites under a zero magnetic field. 
Subsequently, we calculate the isothermal magnetization curves at finite temperatures and investigated the presence of the 1/3 magnetization plateau. Finally, we calculated the adiabatic magnetization curves and the magnetocaloric effect.
Consequently, we estimate the magnetic susceptibility, entropy, and isothermal magnetization curve of the anisotropic triangular lattice of the thermodynamic limit above certain temperatures. 
The 1/3 magnetization plateau at $1\ge J'/J\ge 0.5$ is observed in the isothermal magnetization process. 
In contrast, in the adiabatic process, the anomaly corresponding to the 1/3 magnetization plateau is not flat but inclined.
Regardless of the magnitude of the magnetic field, the magnetization does not reach saturation under the adiabatic process with finite entropy, but the temperature increases rapidly.
Finally, we demonstrate that the magnetic phase boundaries can be determined from the magnetocaloric effect results. 
The results obtained using the OFTLM will be useful for understanding the magnetism of the anisotropic triangular lattice compounds via a comparison with experimental results in the future.

The remainder of this paper is organized as follows. In Sec.~\ref{sec2}, we describe the anisotropic triangular lattice model. 
In Sec.~\ref{sec3}, we describe the FTLM and OFTLM.
In Sec.~\ref{sec4}, we describe the results of the magnetic susceptibility, entropy, isothermal and adiabatic magnetization curves, and  magnetocaloric effect of the anisotropic triangular lattice, and discuss the magnetic properties. 
Finally, a summary is provided in Sec~\ref{sec5}.

\section{Model}
\label{sec2}

The Hamiltonian for the spin-$\frac{1}{2}$ anisotropic triangular lattice in a magnetic field is defined as
\begin{eqnarray} 
\mathcal{H} &=& \sum_{\langle i,j \rangle }J_{i,j} \mathbf{S}_i \cdot \mathbf{S}_j - h\sum_i S^{z}_i,
\label{Hami}
\end{eqnarray}
where $\mathbf{S}_i$ is the spin-$\frac{1}{2}$ operator at the $i$-th site, $S^z_i$ is the $z$ component of $\mathbf{S}_i$, 
$\langle i,j \rangle$ runs over the nearest-neighbor spin pairs, $J_{i,j}$ corresponds to $J$ and $J'$, as shown in Fig.~\ref{lattice}, and $h$ is the magnitude of the magnetic field applied in the $z$ direction.  
Here, we set $J=1$ as the energy unit. 
Notably, in this model, the operator $\sum_i S^{z}_i$ is a conserved quantity because $[\mathcal{H},\sum_i S^{z}_i]=0$.
Here, the eigenvalue of the operator $\sum_i S^{z}_i$ is defined as $S^z_{tot}$.
At $J'=0$, this model becomes the one-dimensional Heisenberg chain, 
whereas at $J'=1$, the model becomes the isotropic triangular lattice.
In the present study, we investigate the model at $J'=0.25, 0.5, 0.75,$ and 1 because several anisotropic triangular lattice compounds have $J\ge J'$~\cite{CCCB,Re2}.

\section{Method}
\label{sec3}
The FTLM has been employed to study the finite-temperature properties of various lattice models~\cite{ftla1,ftla2,ftla3,ftla4,ftla5,ftla6,ftla7,ftla8,ftla9,ftla10,ftla11,ftla12,ftla13}.
The OFTLM is a more accurate method than the standard FTLM, particularly at low temperatures~\cite{OFTL}. 
In this section, we describe the OFTLM and the calculation of physical quantities using this method.
The partition function $Z(T,h)$ of the canonical ensemble at the temperature $T$ in the magnetic field $h$ is expressed as follows:
\begin{eqnarray} 
Z(T,h) = \sum_{m=-M_{\rm sat}}^{M_{\rm sat}}\sum _{i=0}^{N_{st}^{(m)} -1} e^{-\beta E_{i,m}(h)},  \label{Z}
\end{eqnarray}
where $N_{st}^{(m)}$ is the dimension of the Hilbert subspace with $S^z_{tot}=m$ in $\mathcal{H}$, $\beta$ is the inverse temperature 1/$T$ ($k_B=1$), and $E_{i,m}(h)$ is the eigenenergy of the Hilbert subspace with $S^z_{tot}=m$ in $\mathcal{H}$, as a function of $h$.
As $\sum_i S^{z}_i$ is a conserved quantity, 
$E_{i,m}(h)$ is expressed as 
\begin{equation}
E_{i,m}(h) = E_{i,m} - mh, \label{EH} 
\end{equation}
where $E_{i,m}$ is the eigenenergy of the Hilbert subspace with $S^z_{tot}=m$ at $h=0$, and the second term corresponds to the Zeeman term.
We define the order of \{$E_{i,m}$\} as $E_{0,m} \le E_{1,m}  \le E_{2,m} \le \cdots \le E_{N_{st}^{(m)},m} $. 

Using the standard FTLM, the partition function $Z(T,h)$, as shown in Eq.~(\ref{Z}), is approximated as follows:
\begin{equation}
\begin{split}
 Z(T,h)_{\rm FTL} = \sum_{m=-M_{\rm sat}}^{M_{\rm sat}} &\frac{N_{st}^{(m)}}{R}\sum _{r=1}^{R} \sum _{j=0}^{M_L-1}  \\
& e^{-\beta \epsilon^{(r)}_{j,m}(h)} |\langle V_{r,m} | \psi^r_{j,m} \rangle|^2, \label{ZFTL} 
\end{split}
\end{equation}
where $R$ is the number of random samplings of the FTLM, $M_L$ is the dimension of the Krylov subspace,
$|V_{r,m}\rangle$ is a normalized random initial vector with $S^z_{tot}=m$, and $|\psi^r_{j,m}\rangle$ [$\epsilon^{(r)}_{j,m}(h)$] are the eigenvectors (eigenvalues) in the $M_L$-th Krylov subspace with $S^z_{tot}=m$.
Similar to Eq.~(\ref{EH}), $\epsilon^{(r)}_{j,m}(h)$ is expressed as $\epsilon^{(r)}_{j,m}(h)  = \epsilon^{(r)}_{j,m}  - mh$.

In the OFTLM, we first calculate several low-lying exact eigenvectors $| \Psi_{i,m} \rangle$ with $N_V$ levels ($E_{0,m} \le E_{1,m}  \le \cdots \le E_{N_V-1,m} $).
We then calculate the following modulated random vector:
\begin{eqnarray} 
 |V_{r,m}'\rangle  &=& \left[ I - \sum_{i=0}^{N_V-1} | \Psi_{i,m} \rangle \langle \Psi_{i,m} |  \right] | V_{r,m} \rangle,  \label{r'}
\end{eqnarray}
with normalization
\begin{equation}
 |V_{r,m}'\rangle  \Rightarrow \frac{ |V_{r,m}'\rangle }{ \sqrt{\langle V_{r,m}' |V_{r,m}'\rangle} }. \label{r'2}
\end{equation}
Note that $|V_{r,m}'\rangle $ is orthogonal to the states $| \Psi_{i,m} \rangle$ for  $i \in \{ 0, 1, \cdots, N_V-1 \}$.
The partition function of the OFTLM is obtained using $|V_{r,m}'\rangle $ as an initial vector, as follows:
\begin{equation}
\begin{split}
  Z(T,h)_{\rm OFTL} 
 &= \sum_{m=-M_{\rm sat}}^{M_{\rm sat}} \left[ \frac{N_{st}^{(m)}-N_V}{R}\sum_{r=1}^{R} \sum_{j=0}^{M_L-1}  \right. \\
& \left. e^{-\beta \epsilon^{(r)}_{j,m}(h)} |\langle V_{r,m}' | \psi^{r}_{j,m} \rangle|^2  + \sum_{i=0}^{N_V-1}  e^{-\beta E_{i,m}(h)} \right]. \label{ZOFTL} 
\end{split}
\end{equation}
Similarly, in the OFTLM, the energy $E(T,h)_{\rm OFTL}$, magnetization $M(T,h)_{\rm OFTL}$, magnetic susceptibility $\chi(T)_{\rm OFTL}$, and magnetic entropy $S_m(T,h)_{\rm OFTL}$ are obtained as follows:
\begin{equation}
\begin{split}
  E(T,h)_{\rm OFTL}  &= \frac{1}{Z(T,h)_{\rm OFTL}}\sum_{m=-M_{\rm sat}}^{M_{\rm sat}} \left[ \frac{N_{st}^{(m)}-N_V}{R}  \right. \\
      &\times \sum_{r=1}^{R} \sum _{j=0}^{M_L-1} \epsilon^{(r)}_{j,m}(h) e^{-\beta \epsilon^{(r)}_{j,m}(h)} |\langle V_{r,m}' | \psi^{r}_{j,m} \rangle|^2  \\
       &+ \left. \sum_{i=0}^{N_V-1} E_{i,m}(h) e^{-\beta E_{i,m}(h)} \right], 
\label{EOFTL} 
\end{split}
\end{equation}

\begin{equation}
\begin{split}
  M(T,h)_{\rm OFTL}  &= \frac{1}{Z(T,h)_{\rm OFTL}}\sum_{m=-M_{\rm sat}}^{M_{\rm sat}} \left[ \frac{N_{st}^{(m)}-N_V}{R}  \right.  \\
                             &\times \sum _{r=1}^{R} \sum _{j=0}^{M_L-1} m e^{-\beta \epsilon^{(r)}_{j,m}(h)} |\langle V_{r,m}' | \psi^{r}_{j,m} \rangle|^2  \\ 
                             &+ \left.  \sum_{i=0}^{N_V-1} m e^{-\beta E_{i,m}(h)} \right], 
\label{MOFTL} 
\end{split}
\end{equation}

\begin{equation}
\begin{split}
  \chi(T)_{\rm OFTL}  &= \frac{1}{TZ(T,h=0)_{\rm OFTL}}\sum_{m=-M_{\rm sat}}^{M_{\rm sat}} \left[ \frac{N_{st}^{(m)}-N_V}{R} \right.  \\
                             &\times \sum _{r=1}^{R} \sum _{j=0}^{M_L-1} m^2 e^{-\beta \epsilon^{(r)}_{j,m}} |\langle V_{r,m}' | \psi^{r}_{j,m} \rangle|^2  \\
                             &+ \left. \sum_{i=0}^{N_V-1} m^2 e^{-\beta E_{i,m}} \right], 
\label{COFTL} 
\end{split}
\end{equation}
\begin{equation} 
  S_m(T,h)_{\rm OFTL}  = \frac{E(T,h)_{\rm OFTL} }{T} - \ln Z(T,h)_{\rm OFTL}.
\label{SOFTL} 
\end{equation}
The last terms in Eqs.~(\ref{ZOFTL}), (\ref{EOFTL}), (\ref{MOFTL}), and (\ref{COFTL}) are exact values, which are more accurate than those obtained using the standard FTLM, particularly at low temperatures.
Therefore, using the OFTLM, we could evaluate the finite-size effects more accurately.
For subspaces with large $S^z_{tot}$, all the eigenvalues can be calculated using FullED because $N_{st}^{(m)}$ is small.
 Therefore, we use the OFTLM for small $m$ and FullED for large $m$.
The conditions of the calculation are listed in Table~\ref{para27} for $N=27$, Table~\ref{para30} for $N=30$, and Table~\ref{para36} for $N=36$.
We note that $R$, $M_L$, and $N_V$ can be dependent on $m$ in the OFTLM, but we maintain them constant in the present study.
Hereafter, the method combining the OFTLM and FullED is simply called OFTLM for simplicity. 

\begin{table}[tb]
\caption{Conditions of the calculation for the $N=27$ cluster.}
  \begin{tabular}{ c c c c c c } \hline 
    $m$  &  $N_{st}^{(m)}$  & method & $R$ & $M_L$ & $ N_V $ \\ \hline 
   27/2  & 1 & Exact & -- & -- & -- \\ 
   25/2  & 27 & FullED & -- & -- & -- \\ 
   23/2  & 351 & FullED & -- & -- & -- \\ 
   21/2  & 2925 & FullED & -- & -- & -- \\ 
   19/2  & 17550 & FullED & -- & -- & -- \\ 
   17/2  & 80730 & OFTLM & 30 & 100 & 6   \\ 
   15/2  & 296010 & OFTLM & 30 & 100 & 6   \\ 
   13/2  & 888030 & OFTLM & 30 & 100 & 6   \\ 
   11/2  & 2220075 & OFTLM  & 30 & 100 & 6   \\ 
   9/2  & 4686825 & OFTLM  & 30 & 100 & 6   \\ 
   7/2  & 8436285 & OFTLM  & 30 & 100 & 6   \\ 
   5/2  & 13037895 & OFTLM  & 30 & 100 & 6   \\ 
   3/2  & 17383860 & OFTLM  & 30 & 100 & 6   \\ 
   1/2  & 20058300 & OFTLM  & 30 & 100 & 6   \\ \hline 
  \end{tabular}
\label{para27}
\end{table}

\begin{table}[tb]
\caption{Conditions of the calculation for the $N=30$ cluster.}
  \begin{tabular}{ c c c c c c } \hline 
    $m$  &  $N_{st}^{(m)}$  & method & $R$ & $M_L$ & $ N_V $ \\ \hline 
   15  & 1 & Exact & -- & -- & -- \\ 
   14  & 30 & FullED & -- & -- & -- \\ 
   13  & 435 & FullED & -- & -- & -- \\ 
   12  & 4060 & FullED & -- & -- & -- \\ 
   11  & 27405 &FullED & -- & -- & -- \\ 
   10  & 142506 & OFTLM & 30 & 100 & 6   \\ 
   9  & 593775 & OFTLM & 30 & 100 & 6   \\ 
   8  & 2035800 & OFTLM & 30 & 100 & 6   \\ 
   7  & 5852925 & OFTLM  & 30 & 100 & 6   \\ 
   6  & 14307150 & OFTLM  & 30 & 100 & 6   \\ 
   5  & 30045015 & OFTLM  & 30 & 100 & 6   \\ 
   4  & 54627300 & OFTLM  & 30 & 100 & 6   \\ 
   3  & 86493225 & OFTLM  & 30 & 100 & 6   \\ 
   2  & 119759850 & OFTLM  & 30 & 100 & 6   \\ 
   1  & 145422675 & OFTLM  & 30 & 100 & 6   \\  
   0  & 155117520 & OFTLM  & 30 & 100 & 6   \\ \hline 
  \end{tabular}
\label{para30}
\end{table}

\begin{table}[tb]
\caption{Conditions of the calculation for the $N=36$ cluster.}
  \begin{tabular}{ c c c c c c } \hline 
    $m$  &  $N_{st}^{(m)}$  & method & $R$ & $M_L$ & $ N_V $ \\ \hline 
   18  & 1 & Exact & -- & -- & -- \\ 
   17  & 36 & FullED & -- & -- & -- \\ 
   16  & 630 & FullED & -- & -- & -- \\ 
   15  & 7140 & FullED & -- & -- & -- \\ 
   14  & 58905 &FullED & -- & -- & -- \\ 
   13  & 376992 & OFTLM & 10 & 100 & 4   \\ 
   12  & 1947792 & OFTLM & 10 & 100 & 4   \\ 
   11  & 8347680 & OFTLM & 10 & 100 & 4   \\ 
   10  & 30260340 & OFTLM  & 10 & 100 & 4   \\ 
   9  & 94143280 & OFTLM  & 10 & 100 & 4   \\ 
   8  & 254186856 & OFTLM  & 10 & 100 & 4   \\ 
   7  & 600805296 & OFTLM  & 10 & 100 & 4   \\ 
   6  & 1251677700 & OFTLM  & 10 & 100 & 4   \\ 
   5  & 2310789600 & OFTLM  & 10 & 100 & 4   \\ 
   4  & 3796297200 & OFTLM  & 10 & 100 & 4   \\ 
   3  & 5567902560 & OFTLM  & 10 & 100 & 4   \\ 
   2  & 7307872110 & OFTLM  & 10 & 100 & 4   \\  
   1  & 8597496600 & OFTLM  & 10 & 100 & 4   \\ 
   0  & 9075135300 & OFTLM  & 10 & 100 & 4   \\ \hline 
  \end{tabular}
\label{para36}
\end{table}

\begin{figure}[tb]
\includegraphics[width=86mm]{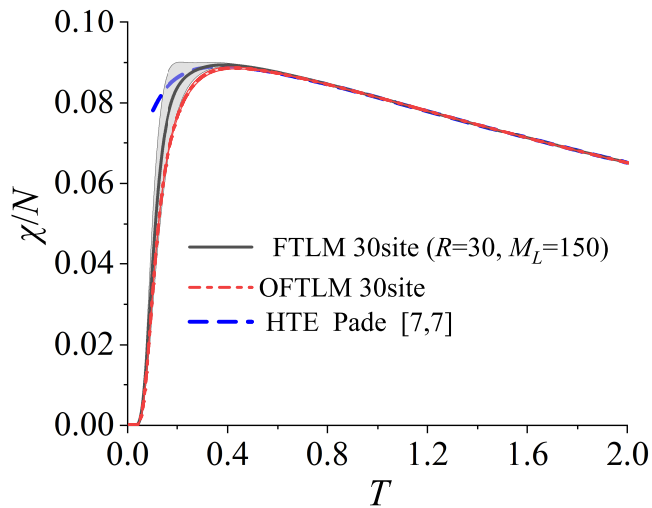}
\caption{
Temperature dependence of the magnetic susceptibility $\chi(T)$ per site for the isotropic triangular lattice ($J'=1$) with $N=30$, obtained using the standard FTLM (black solid line) and OFTLM (red dashed line).
The shaded regions indicate the standard errors of the FTLMs using the jackknife technique.
The result of the high-temperature series expansion combined with the [7,7] Pad\'{e} approximant is also included for comparison (blue dashed line).
\label{Test}}
\end{figure}

\subsection{Benchmark of  the OFTLM}
\label{bench}
We perform benchmark calculations for the standard FTLM and OFTLM. 
We calculate the magnetic susceptibility~$\chi(T)$ per site for the isotropic triangular lattice ($J'=1$) with $N=30$.
The calculation conditions of the standard FTLM are $R=30$ and $M_L=150$.
The calculated results are shown in Fig.~\ref{Test}.
The standard errors of the FTLMs using the jackknife technique~\cite{JK} are represented by the shaded regions.
The result of a previous study on the high-temperature series expansion combined with the [7,7] Pad\'{e} approximant~\cite{THTE} is also included for comparison.
In OFTLM, the error is almost maintained within the line width, whereas standard FTLM has a large error at $T\sim0.2$.
Clearly, the accuracy of the OFTLM is higher than that of the standard FTLM.
In addition, we compare the results of FTLMs with that of the high-temperature series expansion.
As shown in Fig.~\ref{Test}, they are found to be in good agreement for $T>0.4$.
These results suggest that the FTLMs can be used to estimate $\chi(T)$ in the thermodynamic limit with high accuracy, at least for $T > 0.4$.

\begin{figure}[tb]
\includegraphics[width=86mm]{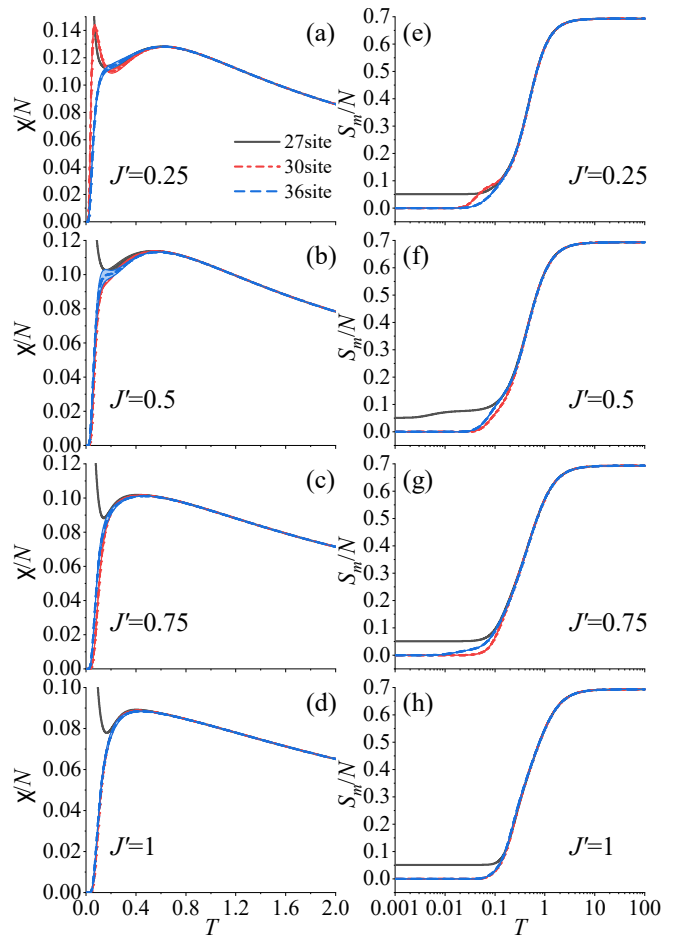}
\caption{
Temperature dependence of the magnetic susceptibility $\chi(T)$ (a--d)
 and magnetic entropy $S_m(T)$ (e--h) per site for the anisotropic triangular lattice with $N=27$, 30, and 36 at $J'=0.25, 0.5, 0.75,$ and 1, obtained using the OFTLM.
The shaded regions indicate the standard errors of the method using the jackknife technique.
\label{chiS}}
\end{figure}

\begin{figure*}[tb]
\includegraphics[width=180mm]{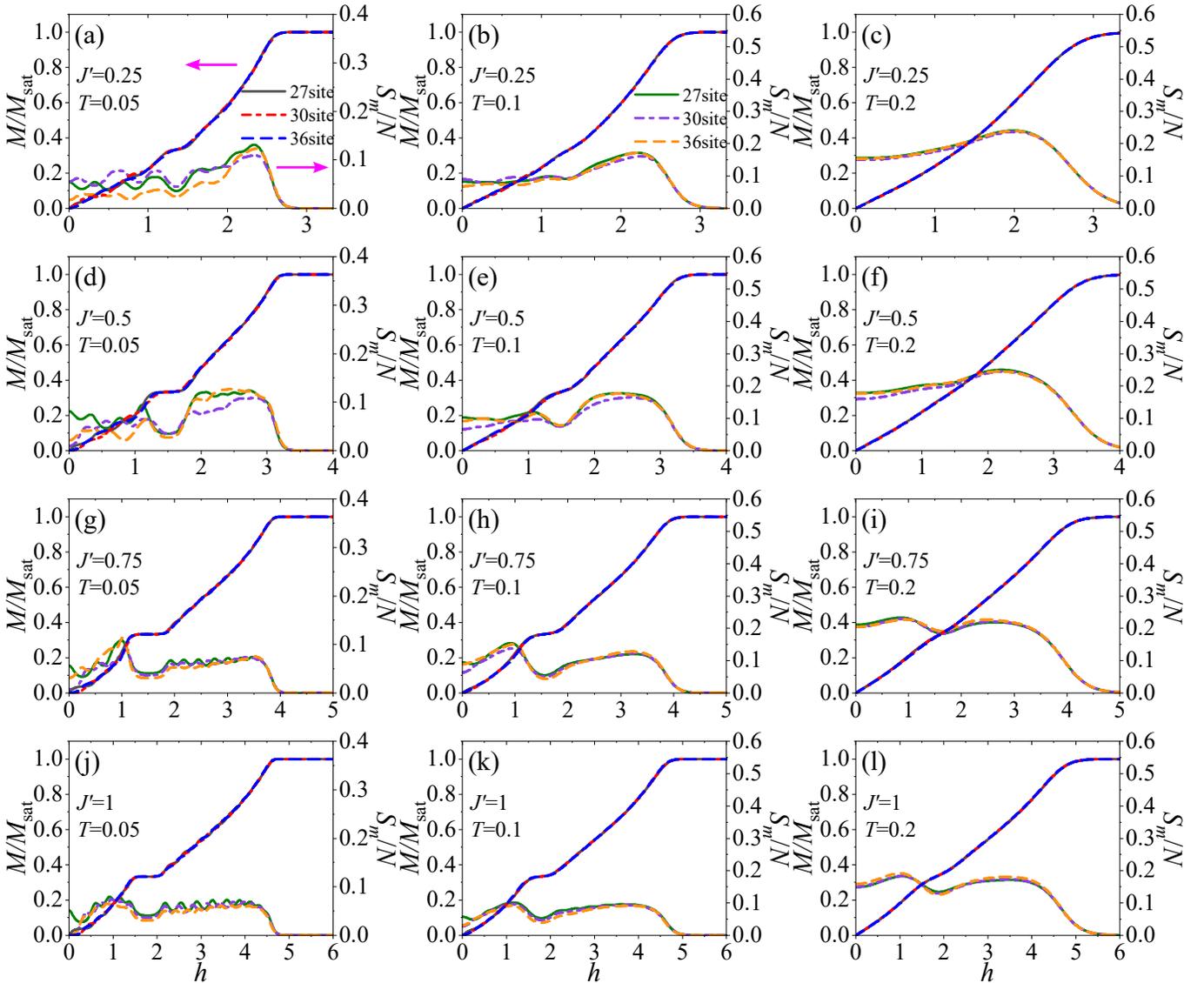}
\caption{Magnetization and entropy curves of the anisotropic triangular lattice with $N=27$, 30, and 36 under the isothermal process at $J'=0.25$~(a--c), $J'=0.5$~(d--f), $J'=0.75$~(g--i), and $J'=1$~(j--l) for $T=0.05$, 0.1, and 0.2, obtained by using the OFTLM.
\label{M-H}}
\end{figure*}

\begin{figure*}[tb]
\includegraphics[width=180mm]{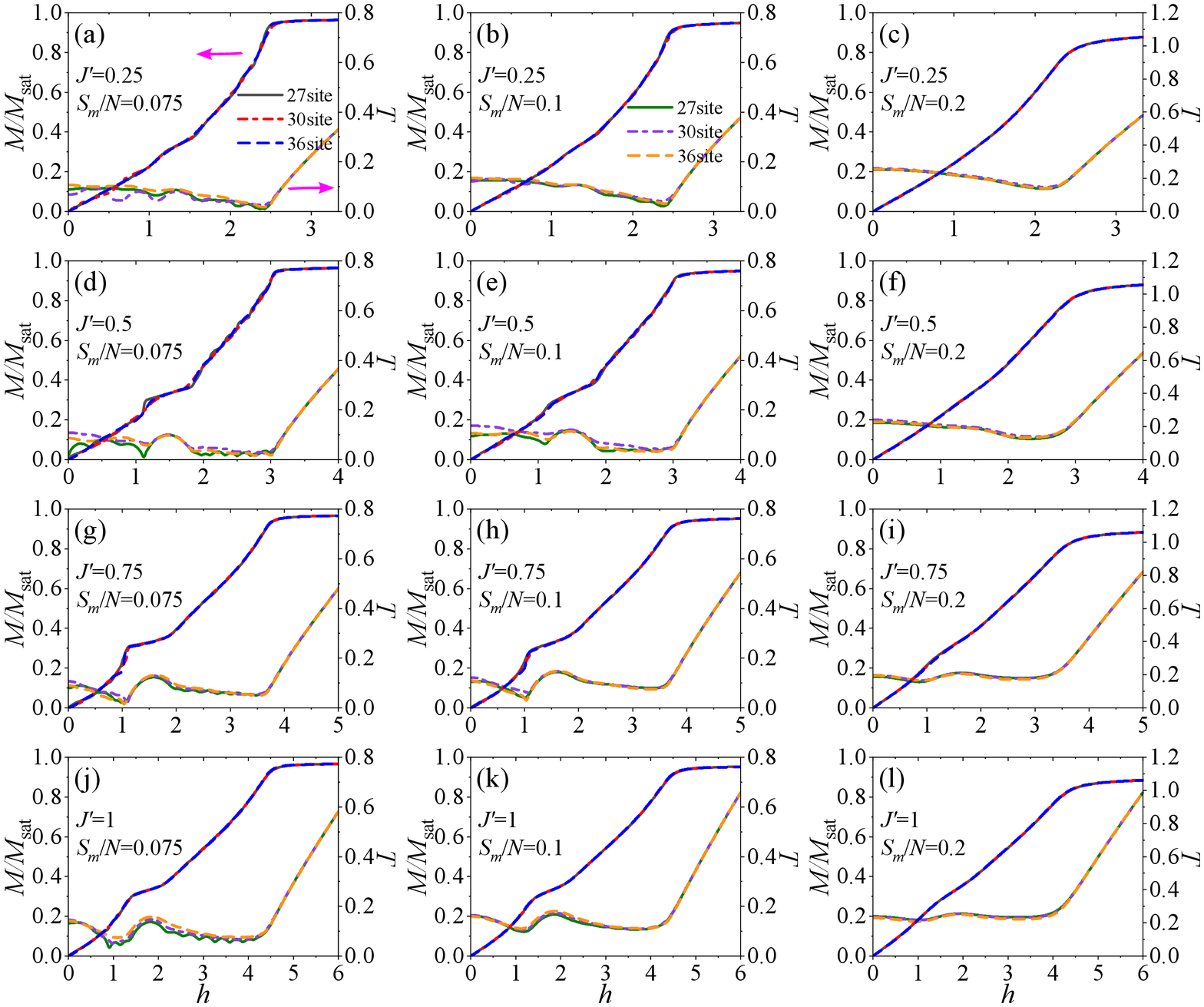}
\caption{Magnetization and temperature curves of the anisotropic triangular lattice with $N=27$, 30, and 36 under the adiabatic process at $J'=0.25$~(a--c), $J'=0.5$~(d--f), $J'=0.75$~(g--i), and $J'=1$~(j--l) for $S_m(T)/N=0.075$, 0.1, and 0.2, obtained by using the OFTLM.
\label{M-H2}}
\end{figure*} 

\section{Results and discussion}
\label{sec4}
\subsection{Magnetic susceptibility and entropy}
\label{subA}
Figure~\ref{chiS} shows the results of the magnetic susceptibility $\chi(T)$ [\ref{chiS}(a), \ref{chiS}(b), \ref{chiS}(c), \ref{chiS}(d)] and magnetic entropy $S_m(T)$ [\ref{chiS}(d), \ref{chiS}(e), \ref{chiS}(f), \ref{chiS}(h)] at $h=0$ for $J'=0.25, 0.5, 0.75$, and 1 at $N=27$, 30, and 36. 
The shaded regions shown in Fig.~\ref{chiS} indicate the standard errors of the OFTLM using the jackknife technique.
In $\chi(T)$, the errors are almost maintained within the line width, whereas in $S_m(T)$, they are sufficiently small compared with the line width.
Therefore, these results make the finite-size effects apparent. 
At $N=27$, in any $J'$, $\chi(T)$ diverges at $T\rightarrow0$, and $S_m(T)$ remains a finite value.
This is because the total magnetization $S^z_{tot}$ of the ground states is not zero, but $\pm1/2$.
$\chi(T)$ at $J'=0.25$, as shown in Fig.~\ref{chiS}(a), has a maximum value at $T\sim0.6$. 
As $J'$ increases, the position of the maximum value decreases to $T\sim0.4$, as shown in Fig.~\ref{chiS}(d).
This displacement of the peak of $\chi(T)$ is consistent with the results of the previous study on the high-temperature series expansion~\cite{ITHTE}.
As shown in Fig.~\ref{chiS}, for $T>0.2$, $\chi(T)$ and $S_m(T)$ are almost independent of size $N$.
Therefore, the positions of the peak of $\chi(T)$ are expected to hardly change, even in the thermodynamic limit. 
Thus, the values of the exchange interactions ($J$ and $J'$) of the model compounds can be estimated by comparing $\chi(T)$ obtained by using the OFTLM and the experimental results for $T>0.2$.
In $S_m(T)$, for $S_m(T)/N>0.1$, almost no size dependence is observed.
This suggests that the adiabatic magnetization process discussed in Sec.~\ref{subC} also has almost no size dependence for $S_m(T)/N>0.1$.

\subsection{Isothermal magnetization process}
\label{subB}
In this subsection, we report the results of the isothermal magnetization process of the anisotropic triangular lattice using the OFTLM.
Figure~\ref{M-H} shows the magnetization and entropy curves for $J'=0.25,$ 0.5, 0.75, and 1 at $N=27$, 30, and 36.
As the numerical errors are small compared with the line width in Fig.~\ref{M-H}, they are not shown.
Here, we first discuss the finite-size effect of the magnetization curve.
At $T=0.2$, almost no size effect is observed, as shown in Figs.~\ref{M-H}(c), \ref{M-H}(f), \ref{M-H}(i), and \ref{M-H}(l).
At $T=0.1$, there is a slight size dependence comparable to the line width at low fields, as shown in Figs.~\ref{M-H}(b), \ref{M-H}(e), and \ref{M-H}(h).
At $T=0.05$, a size dependence exists at low magnetic fields, as shown in Figs.~\ref{M-H}(a), \ref{M-H}(d), \ref{M-H}(g), and \ref{M-H}(j).
These results suggest that the magnetization curve in the thermodynamic limit can be estimated with good accuracy at $T\ge0.1$ using the OFTLM for $N=36$.
From Figs.~\ref{M-H}(e), \ref{M-H}(h), and \ref{M-H}(k), we expect that the 1/3 magnetization plateau exists even in the thermodynamic limit at $T\le0.1$ for $J' \ge 0.5$. 
In addition, in a previous study, at $T=0$, the 1/3 plateau was expected to be observed even at $J' \sim0.3$~\cite{ITLthe1}.
Therefore, we propose that the model compounds of the anisotropic triangular lattice with a very narrow 1/3 plateau or without a 1/3 plateau have $J'<0.5$.
Furthermore, by simultaneously comparing the calculated magnetization curve and susceptibility with those of the model compounds, the exchange interactions ($J$ and $J'$) can be estimated more accurately.

The entropy has minima at the center of the 1/3 plateau as shown in Figs.~\ref{M-H}(d), \ref{M-H}(e), \ref{M-H}(g), \ref{M-H}(h), \ref{M-H}(j), and \ref{M-H}(k). This is due to the fact that the 1/3 plateau state which corresponds to the up-up-down structure with a threefold degeneracy in the thermodynamic limit has an energy gap.
Furthermore, at high magnetic fields, the entropy tends to zero, regardless of $J'$ as shown in Fig.~\ref{M-H}.
This is because there is only one state that has all spins aligned in the magnetic field direction.

\subsection{Adiabatic magnetization process}
\label{subC}
In experiments using pulsed magnetic fields with pulse widths of a few to several tens of microseconds, which have been conducted extensively in recent years~\cite{revew1,revew2}, the magnetization process is not an isothermal process but an adiabatic process because of the very narrow pulse width.
In this subsection, we investigate the adiabatic magnetization process of the anisotropic triangular lattice.

Figure~\ref{M-H2} shows the adiabatic magnetization curves at $N=27$, 30, and 36 using the OFTLM.
The temperature curves under the adiabatic magnetization process, which correspond to the magnetocaloric effect, are also shown.
The magnetization curves and temperature curves were calculated at $S_m/N=0.075$, 0.1, and 0.2.
Here, we first discuss the finite-size effect. 
At $S_m/N = 0.2$, almost no finite-size effect is observed, as shown in Figs.~\ref{M-H2}(c), \ref{M-H2}(f), \ref{M-H2}(i), and \ref{M-H2}(l). 
At $S_m/N = 0.1$, in the magnetization curves, almost no finite-size effect is observed, as shown in Figs.~\ref{M-H2}(b), \ref{M-H2}(e), \ref{M-H2}(h), and \ref{M-H2}(k); however, in the temperature curves, particularly at $J'=0.5$, a size dependence is observed.
At $S_m/N = 0.075$, particularly at $J' \le 0.5$, the size dependence of the magnetization curves and temperature curves is observed.

In the isothermal process, the magnetization curves have a 1/3 plateau for $J'\ge0.5$, whereas in the adiabatic process, the anomaly corresponding to the 1/3 plateau is not flat but inclined, as shown in Figs.~\ref{M-H2}(d), \ref{M-H2}(e), \ref{M-H2}(g), \ref{M-H2}(h), \ref{M-H2}(j), and \ref{M-H2}(k).
The temperature curves have maxima around the center of the region showing this anomaly because the entropy has minima at the center of the plateau in the isothermal process as discussed in Sec~\ref{subB}.
As the temperature is not constant in the adiabatic process, the magnetization curve is not completely flat around $M/M_{\rm sat}=1/3$.
Furthermore, at high magnetic fields, the magnetization $M$ does not reach the saturation magnetization $M_{\rm sat}$, regardless of $J'$.
The entropy at $M=M_{\rm sat}$ is zero as discussed in Sec~\ref{subB}.
In the adiabatic process, as the entropy is constant (non-zero), $M$ never reaches $M_{\rm sat}$, but the temperature increases rapidly, as shown in Fig.~\ref{M-H2}.
Notably, 
at sufficiently small entropy, the adiabatic magnetization curves show the flat 1/3 plateau because the temperature hardly changes during this process.
In the experiment with a high magnetic field and a pulse width of a few microseconds, regardless of the magnitude of the magnetic field, we expect that $M$ does not reach $M_{\rm sat}$ unless the temperature is sufficiently low.

\begin{figure}[tb]
\includegraphics[width=86mm]{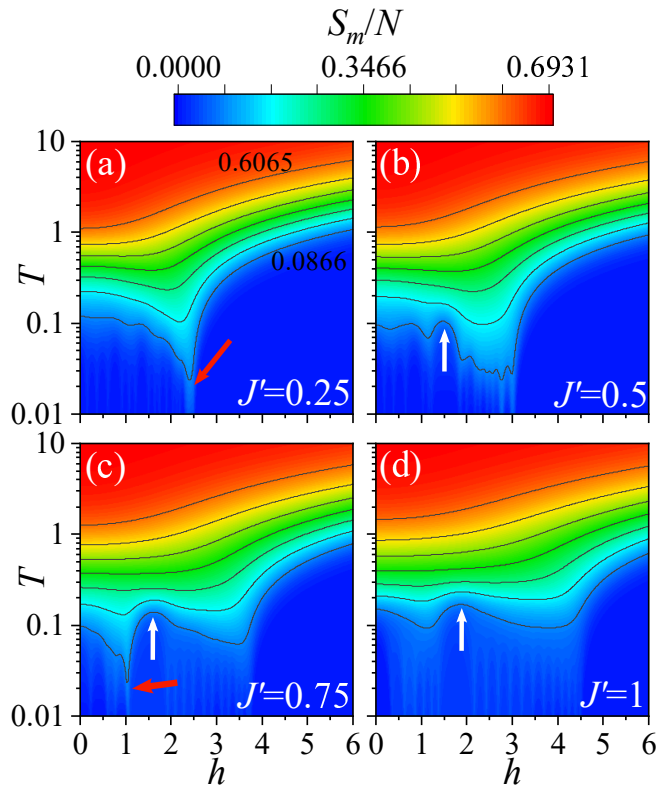}
\caption{Magnetic entropy $S_m$ per site as a function of temperature $T$ and magnetic field $h$ for the anisotropic triangular lattice with $N=36$ using the OFTLM. (a) $J'=0.25$. (b) $J'=0.5$. (c) $J'=0.75$. (d) $J'=1$.
\label{MC}}
\end{figure} 

\subsection{Temperature and magnetic field dependence of the magnetic entropy}
\label{subD}
Figure~\ref{MC} shows the magnetic entropy $S_m$ as a function of temperature $T$ and magnetic field $h$ for the anisotropic triangular lattice with $N=36$ calculated using the OFTLM.
In the low-temperature region $T\le0.05$, vertical streaks are visible owing to the finite-size effect.

For $J'\ge0.5$, the temperature curves at $S_m/N=\frac{1}{8}\ln2 (\sim0.0866)$ have maxima, as indicated by the white arrows in Figs. ~\ref{MC}(b), \ref{MC}(c), and \ref{MC}(d).
These maxima are derived from the 1/3 plateau as described in Sec.~\ref{subC}.
Therefore, such temperature maxima, if experimentally obtained in the magnetocaloric effect measurements, would suggest the presence of a magnetization plateau. 

As shown by the red arrow in Fig.~\ref{MC}(a), 
there is a sharp drop and rise in the temperature under the isentropic process for $S_m/N<0.1$ at $J'=0.25$ around $h=2.5$.
This phenomenon of a sudden temperature change around the critical magnetic field corresponds to the divergence of the magnetic Gr\"{u}neisen ratio $\Gamma_h=\left.\frac{1}{T}\frac{\partial T}{\partial h}\right|_{S_m}$
at a quantum critical point~\cite{MCEthe,ftla4,QCP,QCP2}.
Similarly, at $J '= 0.75$, a rapid temperature change is observed around $h=1.0$, as indicated by the red arrow in Fig.~\ref{MC}(c).
This anomaly would indicate the signature of a quantum phase transition~\cite{ITLthe1}.

We believe that these results can be compared with those obtained experimentally in the future.

\section{Summary}
\label{sec5}
Inspired by the recent development of pulsed magnetic field generators~\cite{revew1,revew2} and the experimental results of anisotropic triangular lattice compounds~\cite{Re1,Re2,Re3}, we investigated the magnetic susceptibility, magnetic entropy, isothermal and adiabatic magnetization curves, and the magnetocaloric effect of the anisotropic triangular lattice using the OFTLM.

We obtained almost size-independent results with $T\ge0.2$ for the magnetic susceptibility and $T\ge0.1$ for the isothermal magnetization curve.
The 1/3 magnetization plateau was observed at $J'\ge0.5$ in the isothermal magnetization process.
By comparing our results for the magnetic susceptibility and isothermal magnetization curve with the experimental results, we could quantitatively determine the exchange interactions ($J$ and $J'$) of the anisotropic triangular lattice compounds.

In the adiabatic magnetization process, the anomaly corresponding to the 1/3 plateau was not flat but inclined.
This is because the entropy of the 1/3 plateau state was lower.
We also obtained the magnetic entropy as a function of the temperature and magnetic field for the anisotropic triangular lattice with $N=36$.
In other words, we obtained the temperature of the adiabatic (isentropic) process as a function of the magnetic field, which corresponds to the magnetocaloric effect. 
We observed an anomaly in the temperature at $J'= 0.75$ around $h=1.0$, which indicates the signature of a quantum phase transition.
We believe that our results will be useful for understanding the experimental results of magnetocaloric effect in the future.

We would like to emphasize that the OFTLM is useful not only for isothermal processes but also for adiabatic processes. 
We hope that our study will motivate further theoretical and experimental investigations of the anisotropic triangular lattice in the future.

\begin{acknowledgments}
We thank Dr. M. Gen and Prof. T. Tohyama for their useful suggestions.
We also thank Prof. T. Tohyama for providing computer resources. 
We would like to thank Editage (www.editage.com) for English language editing.
The numerical calculations were conducted at the facilities of the Supercomputer Center, Institute for Solid State Physics, University of Tokyo.
\end{acknowledgments}

\end{document}